\documentstyle[aps,epsf,prb]{revtex}
\title{Density matrices in $O(N)$ electronic structure calculations: theory 
and applications}
\author{D. R. Bowler\cite{UCL} and M. J. Gillan\cite{DCI}}
\address{Physics and Astronomy Dept., University College London,
Gower Street, London WC1E 6BT, U.K.}
\begin{document}
\twocolumn[\hsize\textwidth\columnwidth\hsize\csname@twocolumnfalse%
\endcsname

\maketitle
\begin{abstract}
We analyze the problem of determining the electronic ground state
within $O(N)$ schemes, focusing on methods in which the total
energy is minimized with respect to the density matrix. We note that
in such methods a crucially important constraint is that the
density matrix must be idempotent (i.e. its eigenvalues must all
be zero or unity). Working within orthogonal tight-binding
theory, we analyze two related methods for imposing this
constraint: the iterative purification strategy of McWeeny,
as modified by Palser and Manolopoulos;
and the minimization technique of Li, Nunes and Vanderbilt. Our analysis
indicates that the two methods have complementary strengths and
weaknesses, and leads us to propose that a hybrid of the two methods
should be more effective than either method by itself. This idea
is tested by using tight-binding theory to apply the 
proposed hybrid method to a set of
condensed matter systems of increasing difficulty, ranging from
bulk crystalline C and Si to liquid Si,
and the effectiveness of the method is
confirmed. The implications of our findings for $O(N)$ implementations
of non-orthogonal tight-binding theory and density functional
theory are discussed. 
\end{abstract}

\bigskip

]

\section{Introduction}
The last few years have seen an upsurge of interest in 
$O(N)$ electronic-structure
methods for treating condensed matter
both within tight-binding theory and within
density functional theory~\cite{(Pettifor 1989),Yang91,(Galli 1992),%
(Li etal 1993),(Daw 1993),(Ordejon etal 1993),(Aoki 1993),mauri93,%
(Goedecker 1994),(Stechel et al 1994),hierse94,kohn95,hernandez95,%
(Kress and Voter 1995),(Horsfield 1996),(Horsfield etal 1996),kohn96,%
hernandez96,goringe97,baer97,haynes97,bowler98}. In these methods,
the number of computer operations needed to determine
the electronic ground state is proportional to the number $N$ of
atoms in the system, instead of showing the $N^2$ or $N^3$
dependence characteristic of traditional methods. $O(N)$
methods are possible because electronic phase coherence is
localised~\cite{kohn95,kohn96,Bowler97}. 
This localisation property can be expressed by saying that
the density matrix $\rho$ decays to zero with increasing distance.

Some practical $O(N)$ methods exploit the locality of
the density matrix directly~\cite{(Li etal 1993),(Daw 1993),kohn95,%
hernandez95,hernandez96,goringe97,haynes97}, 
and have been shown to work particularly 
efficiently for systems with a gap~\cite{Bowler97}.  
They determine the ground state
by minimising the total energy with respect to  $\rho$,
with the approximation that $\rho$ is set equal to zero
for spatial separations exceeding some cut-off $R_{\rm c}$. A central
problem in any such approach is that a density matrix
must be a projector: it is the operator that projects onto
the space of occupied states. Equivalently, one can say that $\rho$
must be idempotent -- its eigenvalues must be zero or
unity. In practice, the requirement of idempotency is
difficult to enforce.

A number of techniques have been suggested for enforcing 
idempotency~\cite{(Li etal 1993),kohn96,McWeeny60,Haynes98,Palser98}.
Many years ago, McWeeny~\cite{McWeeny60} proposed an iterative technique known
as `purification'. The basic idea is that an algorithm is used
to transform a nearly idempotent operator $\tilde{\rho}$ into another
operator $\rho$ that is even more nearly idempotent.
Repetition of this transformation yields a $\rho$ that is idempotent
to any desired precision. Later, Li, Nunes and Vanderbilt 
(LNV)~\cite{(Li etal 1993)} developed a related method using the same 
transformation, in which the total energy is minimised with respect to $\rho$.

The purpose of this paper is to present arguments for
combining these approaches, which we
shall refer to as the McWeeny and LNV approaches. We point
out that the weaknesses of each approach are matched
by the strengths of the other. The fundamental
thought here is that McWeeny is good for finding density
matrices that are idempotent, while LNV is good for
searching through idempotent density matrices to find
the one that yields the true ground state. We shall demonstrate
that this complementarity can be formulated in
a precise and elegant way.

In the next section, we recall the main ideas of the McWeeny and
LNV methods for determining the electronic ground state.
Section~\ref{RelMcWV} is the heart of the paper, where we present our
formulation of the complementarity between McWeeny
and LNV, and we point out its implications
for ground-state search strategy. Practical
illustrations of the benefits obtained by combining the two approaches 
are presented in Sec.~\ref{PracEx}, and in Secs.~\ref{Disc} 
and \ref{Conc} we give discussion and conclusions.

\section{The McWeeny and LNV methods}
\label{McWV}

The arguments we shall develop are very general, but for
simplicity we present them in the framework of orthogonal tight-binding
theory. We comment later on the extensions
to the non-orthogonal case and to density
functional theory.

In the orthogonal tight-binding basis, the matrix elements of the
Hamiltonian are denoted by $H_{i j}$, where $i$ and $j$ go over all
basis functions on all atoms ($1 \le i,j \le N_b$, where $N_b$ is the total
number of basis functions in the system). If there are $N_0$ occupied
states, then the ground-state energy $E_0$ is given by:
\begin{equation}
E_0 = \min {\rm Tr} \,  ( H \rho ) \; ,
\end{equation}
where the minimisation is performed with respect to all Hermitian
matrices $\rho$, subject to the conditions that $\rho$ is
a projector ($\rho^2 = \rho$) and that
${\rm Tr} \, \rho = N_0$. Note that the factor of two due to spin is omitted
for simplicity.  The
ground-state density matrix is given by:
\begin{equation}
\rho = \theta ( \mu I - H ) \; ,
\label{eq:GSRho}
\end{equation}
where $\theta (x)$ is the Heaviside function ($\theta = 1$ for $x > 0$
and $\theta = 0$ for $x < 0$) and $\mu$ is the chemical potential
(Fermi energy). Instead of working at fixed number of occupied states,
it is sometimes more convenient to work at fixed chemical
potential, in which case we minimise the grand potential $\Omega$:
\begin{equation}
\Omega_0 = \min {\rm Tr} \, ( H - \mu I ) \rho
\end{equation}
subject only to the condition $\rho^2 = \rho$.

If one was allowed to diagonalise $H$, then $\rho$ could be
straightforwardly expressed as:
\begin{equation}
\rho_{i j} = \sum_{n = 1}^{N_0} c_{i n} c_{j n}^\ast \; ,
\end{equation}
where $c_{i n}$ are the eigenvector components of $H$:
\begin{equation}
\sum_{j = 1}^{N_b} H_{i j} c_{j n} = \epsilon_n c_{i n} \; ,
\end{equation}
and the eigenvalues $\epsilon_n$ are assumed to be in
ascending order. But diagonalisation is an $O(N^3)$ process and
is incompatible with linear scaling.

\subsection{McWeeny purification}
\label{McWMin}
\begin{figure}
\begin{center}
\epsfxsize=80mm
\epsfbox{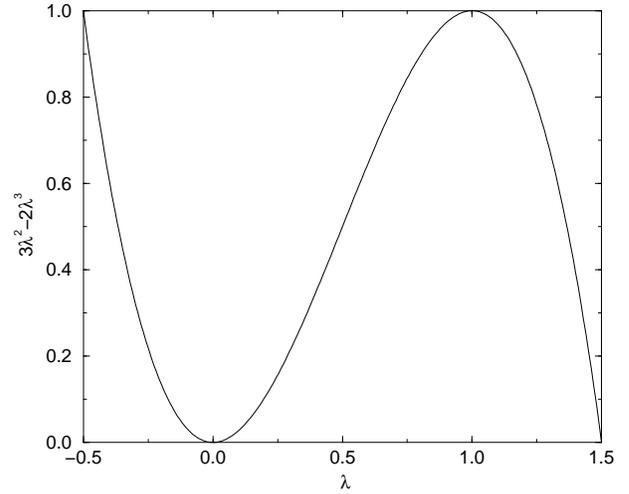}
\end{center}
\caption{The function $f ( \tilde{\lambda} ) = 3 {\tilde{\lambda}}^2
- 2 {\tilde{\lambda}}^3$ in the range $- \frac{1}{2} <
\tilde{\lambda} < \frac{3}{2}$.}
\label{Cubic}
\end{figure}

The McWeeny purification scheme~\cite{McWeeny60} 
gives a way of achieving idempotency
without diagonalisation. The purification algorithm for mapping
a nearly idempotent matrix $\tilde{\rho}$ into one that is more nearly
idempotent is:
\begin{equation}
\rho = 3 {\tilde{\rho}}^2 - 2 {\tilde{\rho}}^3 \; .
\label{eq:McW}
\end{equation}
The way this mapping works can be understood by considering the eigenvalues
of $\rho$ and $\tilde{\rho}$, denoted by $\lambda$, $\tilde{\lambda}$
respectively. Since $\rho$ is diagonal in any representation
that diagonalises $\tilde{\rho}$, the relationship between their
eigenvalues is:
\begin{equation}
\lambda = 3 {\tilde{\lambda}}^2 - 2 {\tilde{\lambda}}^3 \; .
\end{equation}
From the form of the function $f ( \tilde{\lambda} ) =
3 {\tilde{\lambda}}^2 -2 {\tilde{\lambda}}^3$ (see Figure~\ref{Cubic}),
it follows that if $- \frac{1}{2}
< \tilde{\lambda} < \frac{3}{2}$ then $0 \le \lambda \le 1$.
Furthermore if $0 < \tilde{\lambda} < 1$ then
for $\tilde{\lambda} < \frac{1}{2}$,
$\lambda < \tilde{\lambda}$ and for $\tilde{\lambda} > \frac{1}{2}$,
$\lambda > \tilde{\lambda}$, so that iteration of the
mapping drives the eigenvalues towards the values 0 or 1. As
this process continues, the final approach to idempotency
accelerates rapidly. If $\tilde{\lambda}$ deviates
from zero by a small amount,
then the deviation of $\lambda$
is proportional to the square of that amount, and
similarly for deviations from unity.
This quadratic convergence to idempotency can be summarised
by noting that the matrices $\rho^2 - \rho$ and
${\tilde{\rho}}^2 - \tilde{\rho}$ are related by:
\begin{equation}
\rho^2 - \rho = 4 ( {\tilde{\rho}}^2 - \tilde{\rho} )^3 -
3 ( {\tilde{\rho}}^2 - \tilde{\rho} )^2 \; .
\end{equation}

For a given initial matrix $\rho^{(0)}$, iteration of
the purification algorithm therefore generates a sequence
$\rho^{(1)}$, $\rho^{(2)}$,... that converges to an idempotent matrix
$\rho^{( \infty )}$. However, there are many idempotent matrices,
and $\rho^{( \infty )}$ is not necessarily the idempotent
matrix given by eqn~(\ref{eq:GSRho}). The latter is uniquely
specified by the statements that (a) $\rho$ commutes with $H$;
(b)~in a representation that diagonalises $\rho$ and $H$, the
eigenvalues $\lambda_n$ of $\rho$ (which by idempotency
must be 0 or 1) are given by $\lambda_n = 1$
for $\epsilon_n < \mu$ and $\lambda_n = 0$ for
$\epsilon_n > \mu$.

It has been emphasised recently by Palser and Manolopoulos~\cite{Palser98}
(hereafter PM) that McWeeny purification automatically
delivers the correct ground state provided the initial
$\rho$ is an appropriate function of the Hamiltonian whose
eigenvalues are in the range $(0,1)$.
The initial $\rho$ then commutes with $H$, and this means
that all subsequent $\rho^{(k)}$ commute with $H$. Provided
the eigenvalues of $\rho^{(0)}$ satisfy $0 < \lambda_n < \frac{1}{2}$
for $\epsilon_n > \mu$ and $1 > \lambda_n > \frac{1}{2}$
for $\epsilon_n < \mu$, then repeated purification
automatically yields a final $\rho$ representing the exact
ground state. PM point out that $\rho^{(0)}$ satisfies
the requirements if it is chosen as:
\begin{equation}
\rho^{(0)} = \frac{1}{2} \xi ( \mu I - H ) +
\frac{1}{2} I \; ,
\label{eq:Rho0}
\end{equation}
where
\begin{equation}
\xi = \min \left\{ \frac{1}{H_{\rm max} - \mu} ,
\frac{1}{\mu - H_{\rm min}} \right\} \; ,
\label{eq:HamLim}
\end{equation}
with $H_{\rm min}$, $H_{\rm max}$ lower and upper bounds on
the eigenvalue spectrum of $H$. A crucial feature of this
procedure is that the grand potential $\Omega^{(n)}$ converges monotonically
to the ground state value from above: $\Omega^{(n+1)} <
\Omega^{(n)}$. PM further show that a purification
procedure working at fixed $N_0$ can be obtained with
a modified version of the McWeeny algorithm;
details are given in
their paper. In this case, the energy $E^{(n)}$ converges
monotonically from above: $E^{(n+1)} < E^{(n)}$.

The McWeeny purification scheme, as modified by PM, should therefore 
provide an extremely effective method for determining the ground state,
provided everything is done exactly. But the essence of
$O(N)$ methods is the imposition of a spatial cut-off
on the density matrix. This means that at each purification
step the input matrix ${\tilde{\rho}}_{i j}$ will be non-zero
only if $j$ is one of a spatially localised set of neighbours
of $i$. Because of the matrix multiplications, the output
matrix $\rho_{i j}$ will have a more extended range, so that
it must be truncated back to the imposed range before each
iteration. Because of this truncation, the monotonic
convergence of $\Omega$ or $E$ must fail as the ground state
is approached, and PM suggest that this failure of monotonicity
be used as a criterion for terminating the iterative process.
In other words, should the energy for any given iteration be higher
than that from the previous iteration, the process should be stopped. 
This is an important criterion to follow, as once the truncation errors 
become significant, the iteration will no longer converge on an idempotent
matrix.

This heuristic criterion for terminating
the iterations may work in some cases,
but we consider it to be unsatisfactory for two reasons.
First, it makes the approximate ground-state energy depend
on the details of the initial $\rho^{(0)}$. Second, the energy
is not the minimum of any function, so that the variational
property of the exact ground state is lost. This means that
calculated forces on atoms will not be consistent with
the energy, so that both relaxation to equilibrium and dynamics
are likely to be problematic. In addition, the use of this
procedure as part of an $O(N)$ density functional scheme
would encounter other problems, as we point out in Sec.~\ref{Disc}.

\subsection{LNV minimisation}
\label{VandMin}

The scheme of Li, Nunes and Vanderbilt~\cite{(Li etal 1993)} 
also makes use of the purification algorithm,
but in a completely different way. If we work at constant
$\mu$, then the strategy is based on minimisation of
$\Omega = {\rm Tr} \, ( H - \mu I ) \rho$,
subject to the condition that $\rho$ is weakly idempotent (this
means that its eigenvalues $\lambda$ satisfy $0 \le \lambda \le 1$).
In this domain of matrices, the minimum value of $\Omega$ is
obtained when $\rho$ is given by eqn~(\ref{eq:GSRho}). This is easily seen
by writing $\Omega$ in a representation in which $H$ is
diagonal (but $\rho$ is not assumed to be diagonal):
\begin{equation}
\Omega = \sum_n ( \epsilon_n - \mu ) \rho_{n n} \; .
\end{equation}
The diagonal elements of a weakly idempotent matrix must lie
in the range $[0,1]$, so that the minimum of $\Omega$ is obtained
when $\rho_{n n} = 1$ for $\epsilon_n < \mu$ and
$\rho_{n n} = 0$ for $\epsilon_n > \mu$. In this case,
it is readily shown that all the
off-diagonal elements of $\rho_{n n}$ must vanish, and $\rho$
is given by eqn~(\ref{eq:GSRho}).

The constraint of weak idempotency can be achieved by expressing
$\rho$ as in eqn~(\ref{eq:McW}), provided the eigenvalues of
$\tilde{\rho}$ lie in the range $( - \frac{1}{2} , \frac{3}{2} )$.
In this role, $\tilde{\rho}$ is simply an auxiliary matrix,
introduced solely to satisfy weak idempotency. Then $\Omega$
can be written as:
\begin{equation}
\Omega = {\rm Tr} \, ( H - \mu I )
( 3 {\tilde{\rho}}^2 - 2 {\tilde{\rho}}^3 ) \; .
\label{eq:GP}
\end{equation}
But since this is a cubic form in $\tilde{\rho}$, it can have only
a single minimum, which is obtained when $\tilde{\rho} =
\rho = \theta ( \mu I - H )$. In practice, it is often more
convenient to minimise the energy $E = {\rm Tr} \,
H ( 3 {\tilde{\rho}}^2 - 2 {\tilde{\rho}}^3 )$ subject to the
constraint that the number of occupied states $N_0 =
{\rm Tr} \, ( 3 {\tilde{\rho}}^2 - 2 {\tilde{\rho}}^3 )$ is
held constant.  This requires more computational effort, since the gradient of 
the electron number with respect to the density matrix must be evaluated; 
there are several efficient implementations of 
the constant-$N_0$ constraint~\cite{CMG,KMHo}.

The great advantage of LNV over McWeeny is that it is variational,
and this means that it works even when a spatial cut-off is imposed on
$\tilde{\rho}$. Indeed, it is one of the standard methods of
achieving linear-scaling behaviour in tight-binding
calculations~\cite{Bowler97}. With a cut-off, the minimum
of $\Omega$ or $E$ is guaranteed to be above the exact ground-state
value, and this minimum decreases monotonically to the exact
value as the cut-off is increased. The variational property means
that forces on the atoms calculated at the minimum are exactly
consistent with the variations of $\Omega$ or $E$.  

Nevertheless, the LNV techique does have several weaknesses, all
of which affect the process of searching for the ground state.
An obvious weakness is that it does not have the quadratic
convergence shown by the McWeeny technique. Any iterative
method used to minimise $\Omega$ or $E$ will give linear convergence,
so that as we approach the minimum the error in $\tilde{\rho}$
at each iteration is some fraction of the error at the previous
iteration. This means that -- at least in the absence of
a spatial cut-off -- LNV is expected to need more iterations
than McWeeny to achieve a given accuracy.

This weakness is exacerbated by the fact that the LNV technique
demands more operations in each iteration, as has been emphasised
by PM~\cite{Palser98}. This is simply because
we need to calculate the gradient of $\Omega$ or $E$
with respect to the elements of $\tilde{\rho}$,
in addition to calculating $\Omega$ or $E$ itself. So even if
it did not need more iterations, LNV would still be slower.
Roughly speaking, an LNV iteration takes about twice as 
long as a McWeeny iteration in constant-$N_0$ calculations
and somewhat more than this in constant-$\mu$ calculations.

There is also a third cause of slowness in LNV, namely ill conditioning.
In any minimisation problem, convergence to the minimum
will be slow if the curvatures of the function are very different
in different directions, or equivalently if the eigenvalues
of the Hessian matrix span a wide range. But it is readily shown
that the curvatures of $\Omega$ are determined by the
quantities $\mid \epsilon_n - \mu \mid$,
which will indeed span a wide range unless
the system has a large band-gap. We should expect this weakness to
be particularly troublesome for metallic systems, and in fact detailed
evidence for the inefficiency of LNV for such systems
has already been presented~\cite{Bowler97}. The McWeeny technique
does not suffer from this problem.

In addition to these problems of convergence speed, LNV has two other 
obstacles: initialisation and poor robustness.  It is far from clear 
how best to initialise the density matrix within LNV; in general, the 
{\it ansatz} of $\tilde{\rho} = \frac{1}{2}{\bf I}$ is made within orthogonal
tight binding\cite{(Li etal 1993),CMG}.  Poor robustness arises 
since $\Omega$ is a cubic form in
$\tilde{\rho}$ and is unbounded below. In order for a minimisation
method to lead to the minimum of $\Omega$, the initial
$\tilde{\rho}$ must be chosen close enough to the minimum. If we start
from an unsuitable initial $\tilde{\rho}$, any downhill search method
will lead away from the minimum, and $\Omega$ will plunge
towards infinitely negative values. In a robust search strategy
we expect to seek the minimum of $\Omega$ in a sequence of
search directions. It is a sign of danger if $\Omega$ has a point of
inflection but no minimum in a search direction,  
and we shall use this idea later 
when discussing robustness.

We show in the next section how a combination of McWeeny and LNV
allows their strengths to be exploited and their weaknesses
to be avoided.

\section{Relationship between McWeeny and LNV}
\label{RelMcWV}

In analysing the relationship between the two methods, we find it
helpful to regard matrices as elements of a vector space. Using
this viewpoint, we shall make free use of geometrical
concepts such as the straight line joining two matrices $A$
and $B$, by which we mean the set of matrices $( 1 - \lambda ) A +
\lambda B$, where $0 < \lambda < 1$. We define the magnitude or
norm of a matrix $A$ as:
\begin{equation}
\mid\mid A \mid \mid = [ {\rm Tr} \, ( A^\dagger A ) ]^{1/2} \; ,
\label{eq:Norm}
\end{equation}
where $A^\dagger$ is the Hermitian conjugate of $A$. This allows
us to talk of the `distance' $\mid \mid A - B \mid \mid$ between
two matrices. We shall also need the notion of scalar product
of two matrices, defined as:
\begin{equation}
( A , B ) = {\rm Tr} \, ( A^\dagger B ) \; .
\label{eq:AB}
\end{equation}
Matrices are referred to as `orthogonal' if $( A , B ) = 0$.

In the vector space just defined, there is a manifold consisting
of all matrices that are idempotent, and we call this the
idempotency surface. For any suitably chosen matrix $A^{(0)}$,
the purification algorithm (Eq.~(\ref{eq:McW})) generates a sequence of
matrices $A^{(1)}$, $A^{(2)}, ...$ which tend to a limit
$A^{(\infty)}$ lying on the idempotency surface. If we imagine
this sequence of points in the vector space as joined by
straight lines, then we form a path which we refer to as
the McWeeny path.

\begin{figure}
\begin{center}
\epsfxsize=80mm
\epsfbox{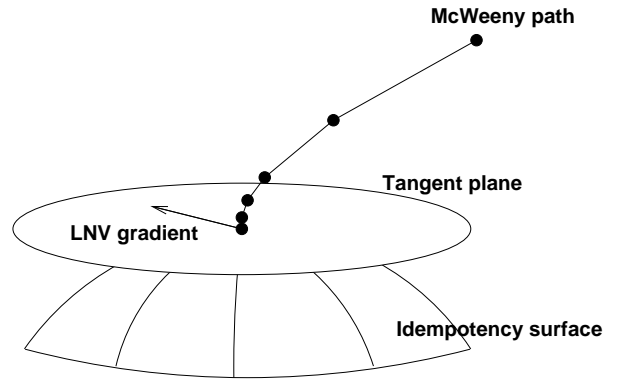}
\end{center}
\caption{Geometrical representation of the McWeeny
and LNV methods, showing that the density matrices generated by
McWeeny iteration form a path approaching the idempotency
surface orthogonally, and that the gradient of the LNV
function lies in the tangent plane to the idempotency surface.}
\label{IdemSurf}
\end{figure}

The following two statements, proved in Appendix~\ref{AppOrtho}, will
play a key role in our proposed strategy for finding the
ground state:

\begin{enumerate}
\item
All McWeeny paths meet the idempotency surface orthogonally;

\item
For any point on the idempotency surface, the gradient of the
LNV function is tangential to the surface.
\end{enumerate}
A picture illustrating these statements is shown in Fig.~\ref{IdemSurf}.

To explain more precisely what these statements mean,
we need to define the concept of tangent planes to the
idempotency surface. For any point $P$ on this
surface ($P^2 = P$), consider points on the straight line
$\tilde{P} = P + \alpha B$, where $B$ is some chosen
Hermitian matrix and $\alpha$ is a real scalar variable.
In general, $\tilde{P}$ will not be idempotent,
but for some choices of $B$, $\tilde{P}$ is idempotent
to linear order in $\alpha$:
\begin{equation}
{\tilde{P}}^2 - \tilde{P} = \alpha^2 B^2 \; .
\end{equation}
For such choices of $B$, the shortest distance between a given
point on the straight line and the idempotency surface
is of order $\alpha^2$, and 
we can say that the straight line is a tangent
line to the idempotency surface. The tangent plane at the point
$P$ consists of all matrices $\tilde{P}$ on all tangent lines
passing through $P$. A convenient way to construct points
in the tangent plane is described in Appendix~\ref{AppOrtho}.

Now suppose we have a McWeeny sequence $A^{(k)}$ going to the
limit $A^{(\infty)}$. Then the meaning of statement~(1)
is that in the $k \rightarrow \infty$ limit the difference
vector $A^{(k)} - A^{( \infty )}$ becomes orthogonal
to every vector $B$ for which $A^{( \infty )} + B$ is in
the tangent plane passing through $A^{( \infty )}$:
\begin{equation}
\lim_{k \rightarrow \infty} ( B , A^{(k)} - A^{( \infty )} )
\left/ \mid \mid A^{(k)} - A^{( \infty )} \mid \mid \right. = 0 \; .
\end{equation}
The meaning of statement~(2) is that, if $F$ is the gradient of
the LNV function at point $P$ on the idempotency surface,
then $P + \alpha F$ is in the tangent plane passing
through $P$. The corollary is that for idempotent $P$ the LNV
gradient is orthogonal to the McWeeny path.

Consider the implications of the two statements. We know that
purification is a completely robust way of reaching idempotency.  Statement~(1)
implies that it is also very direct. As we approach the idempotency
surface, we are following the shortest possible path. Once
we are near the surface, purification is tantamount to
dropping a perpendicular onto the surface. In addition, quadratic
convergence means that the approach to the surface accelerates
rapidly in the final stages. But for a general starting point
purification does not give us the ground state. It gives us {\em an}
idempotent density matrix, but not {\em the} idempotent density
matrix corresponding to the ground state. This is where LNV
comes in. Once we are on the idempotency surface, the two statements
guarantee that application of LNV does not undo what
was achieved by McWeeny, because it keeps us on the surface
to first order. Furthermore, LNV is a completely
robust way of finding the ground state if we are constrained to
the surface. Within the idempotency constraint, the LNV
function has only a single minimum, and cannot fall below the
ground state energy.

We have assumed up to now that everything is done exactly,
without any spatial cut-off. In this case, purification
by itself is enough to find the ground state. As stressed
by PM (see Sec.~\ref{McWV}), an initial guess for $\rho^{(0)}$ given
by eqns~(\ref{eq:Rho0}) and (\ref{eq:HamLim}) 
guarantees that repeated purification
delivers the unique density matrix corresponding to the
ground state. But with a cut-off the situation is quite
different. Purification then brings us near to the ground state,
but leaves us with a ground-state estimate lacking
variational properties. At this point, we suggest that the
effective strategy is to switch to LNV. Since we are
already near the ground state, minimisation of the LNV
function should be rapid. Although we are not exactly
on the idempotency surface, the energy gradient should still
maintain idempotency to good accuracy, and there should
be no danger of approaching the unstable region where
the LNV function decreases unboundedly.

An important benefit of combining McWeeny and LNV in the way we suggest
is that it makes it easier to work at constant electron number, which
is usually what one wishes to do. Although the modifications proposed
by PM make it straightforward to perform McWeeny purification
at constant $N_0$, the minimization of the LNV function is made more 
complicated by the need to hold $N_0$ fixed. But once we are on the
idempotency surface, this difficulty in the LNV scheme disappears.
This can be seen by recalling that $N_0 = {\rm Tr} \,
[ 3 {\tilde{\rho}}^2 - 2 {\tilde{\rho}}^3 ]$, so that
$\partial N_0 / \partial {\tilde{\rho}}_{i j} =
6 ( \tilde{\rho} - {\tilde{\rho}}^2 )_{i j}$, which vanishes
if $\tilde{\rho}$ is idempotent. The implication is that, since
the LNV gradient keeps $\tilde{\rho}$ near the idempotency
surface once it has been brought there by McWeeny purification,
the electron number will automatically maintain itself almost
constant during the LNV stage. Practical tests of this will be
shown in the next section.

In summary, the proposed hybrid strategy capitalizes on the expected
robustness and speed of purification, but avoids its lack of
variational properties. Variational behaviour is supplied
by the LNV component of the strategy, but since this
is used only in the final stages, we hope to avoid the instability
and slowness from which LNV can suffer if used alone.
An important feature to note is that, as the cut-off radius increases,
we expect the hybrid strategy to become increasingly
effective: almost all the work is then done by
purification, so that the strengths of McWeeny and LNV
should complement each other better and better.

\section{Practical examples}
\label{PracEx}

In this section we present examples of calculations done using the
proposed hybrid
approach, compared with pure LNV calculations; these examples form a 
sequence of systems of increasing difficulty: bulk, diamond-structure carbon
and silicon; a relaxed carbon vacancy; the Si(001) surface; and finally 
liquid silicon.

There are two main characteristics that we 
are looking for in these tests: speed 
and robustness.  
The speed of a method can be gauged either by the total 
number of iterations 
taken, or by the total CPU time used.  These give different
measures, since each 
McWeeny iteration requires only about half as much time as an 
LNV iteration. When comparing different methods, we shall characterize
the speed by the number of iterations needed to reach the
ground state within a specified tolerance, and the implications
for CPU time will be pointed out where appropriate.

Robustness is harder to characterize. As explained in Sec.~\ref{VandMin},
the LNV method can become unstable if the grand potential
$\Omega$ has no minimum in a search direction, because this signals that we
are approaching a dangerous region where $\Omega$ decreases unboundedly.
This behaviour is, indeed, found on occasions in calculations based on
pure LNV, particularly when the electron number is kept constant (as opposed
to the electron chemical potential). The kind of robustness we are 
looking for in our hybrid method is therefore the absence of this kind of 
instability.  Rather than repeat the refrain throughout each section, we 
will state now that we have not seen this behaviour at any time during 
these simulations.

To perform the tests, we have used a simple, 
nearest-neighbour orthogonal tight binding model, with parameterisations for 
silicon~\cite{DRB} and carbon~\cite{APH}.  We adapted an implementation of 
the LNV scheme by Goringe~\cite{CMG}.  In this scheme, the density matrix 
cut-off is not defined as a sphere, but rather in terms of a cluster of atoms.
This cluster is formed by including all atoms which are within range of a 
certain number of `hops' of the central atom (e.g. a nearest neighbour is at 
one hop, and its nearest neighbours are at two hops from the central 
atom).  The localisation criterion is thus specified
in terms of the number of hops.  All McWeeny minimisations have been carried 
out at fixed electron number; except where stated, the same is true for LNV 
minimisations and LNV stages of hybrid minimisations.

The initial density matrix for the pure LNV method was chosen to be
$\frac{1}{2}${\bf I}, while for the hybrid and McWeeny methods it was 
constructed according to equations~(\ref{eq:Rho0}) and (\ref{eq:HamLim}).
When using the hybrid method, the switch to the LNV phase from the McWeeny
phase occurred when the energy in one iteration was higher than that in the
previous iteration (which is indicative of truncation error) or an equivalent
error was found in the maintenance of electron number (as described by Palser
and Manolopoulos\cite{Palser98}).  As the density matrix had been shown to be
invalid, the density matrix from the {\it previous} iteration was passed 
to the LNV phase as an initial matrix.

\subsection{Perfect Si and C crystals}

We have used our proposed hybrid scheme and the pure LNV 
scheme to find the electronic
ground state for carbon and silicon in the diamond crystal structure for 
different cut-off radii of the density matrix (3, 5 and 7 hops).
Tables~\ref{Tab:CConv} and~\ref{Tab:SiConv} give the number 
of iterations taken by the two methods to achieve a specified tolerance in 
fractional change in cohesive energy between line minimisations (in this case, 
$10^{-8}$).  This is a standard type of criterion applied in practical 
minimisation, and is used here to show the performance of the methods in 
practical tests; in both cases, the methods are achieving the ground state, 
which ensures that a fair comparison is made.  
For the hybrid scheme, we give separately the 
numbers of iterations needed in the McWeeny and LNV stages of
the calculation, while for the LNV scheme we simply give the
total number of iterations.  As described above in sections \ref{McWMin} and
\ref{VandMin}, the initialisations used are the PM\cite{Palser98} $\rho^{(0)}$
for the hybrid initialisation and $\frac{1}{2}{\bf I}$ for the LNV 
initialisation.  Several points should
be noted. First, the total numbers of iterations are the essentially
the same for both schemes. As pointed out above, this means that
the hybrid scheme is significantly faster (by about 25 \%).
Second, the LNV stage of the hybrid scheme takes fewer iterations
as the cut-off radius is extended. This is expected, because
the McWeeny stage should bring the density matrix closer to
the ground state for larger radii. To confirm this point, we
have calculated the norm (see Eq.~(\ref{eq:Norm})) of the difference
between the density matrix obtained
at the end of the McWeeny stage and
final ground-state density matrix. This norm, reported in the
last column of Tables~\ref{Tab:CConv} and \ref{Tab:SiConv},
decreases markedly with increasing cut-off radius. Finally, it is
worth noting that the Si crystal takes somewhat more iterations that
the C crystal, as might be expected because of its smaller band gap.

\begin{figure}
\begin{center}
\epsfxsize=80mm
\epsfbox{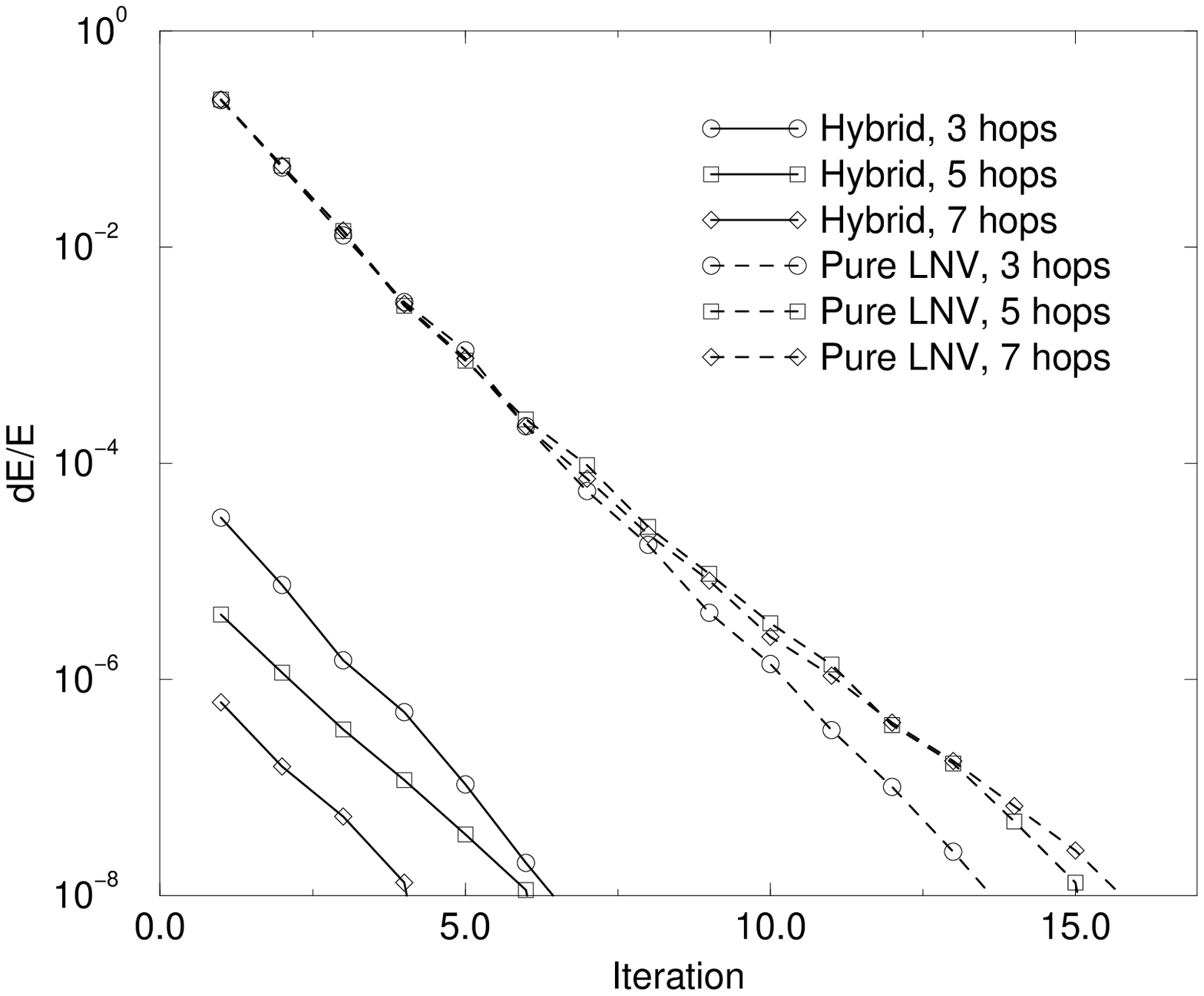}
\end{center}
\caption{The difference between the cohesive energy at a given iteration and
the final cohesive energy for the LNV stage of the hybrid scheme (solid lines)
and the pure LNV scheme (dashed lines) for diamond structure carbon.  
Results are shown for different cut-off radii: 3 hops (circles), 
5 hops (squares) and 7 hops (diamonds).  The radii are discussed in the text.}
\label{CConv}
\end{figure}

\begin{figure}
\begin{center}
\epsfxsize=80mm
\epsfbox{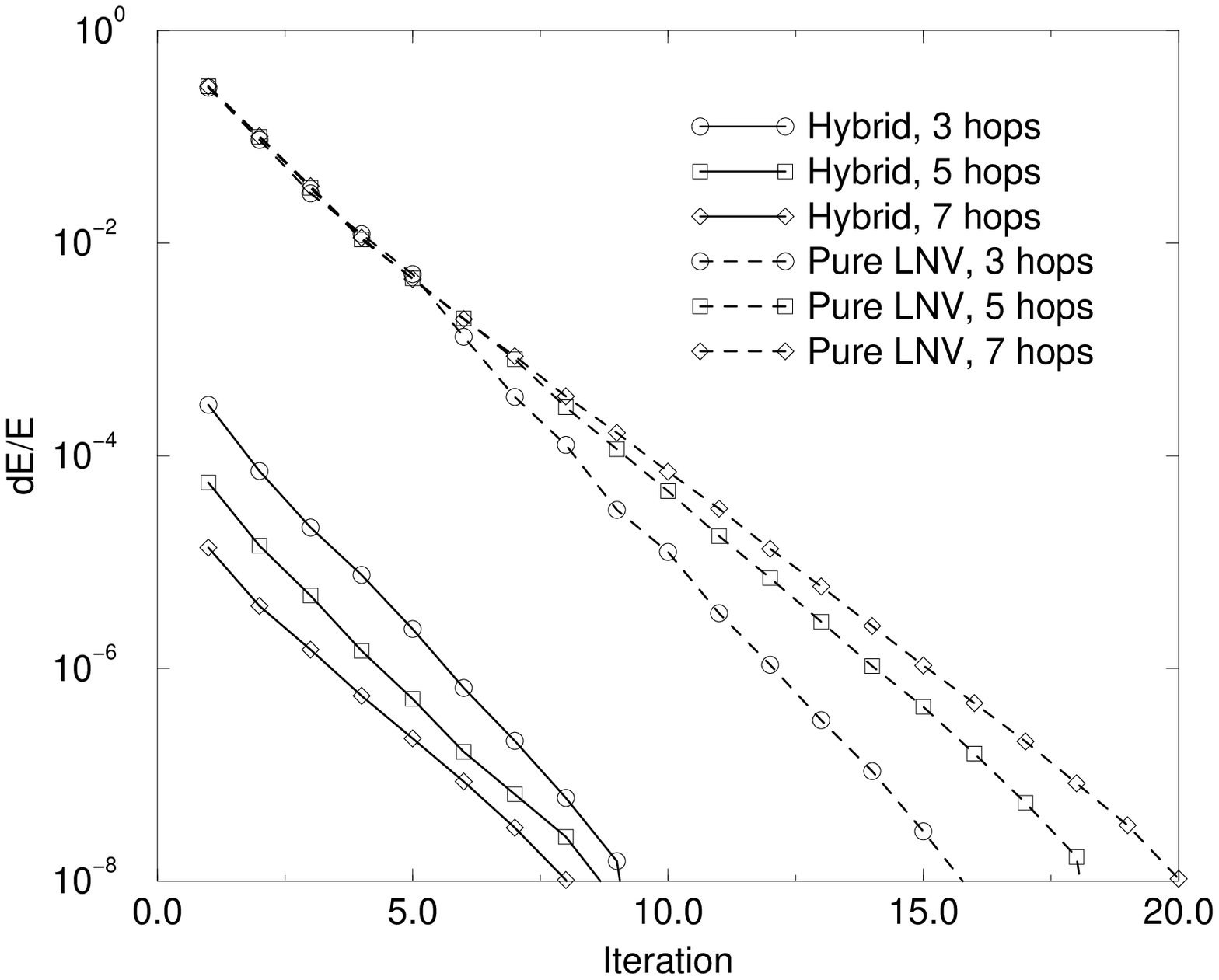}
\end{center}
\caption{The difference between the cohesive energy at a given iteration and
the final cohesive energy for the LNV stage of the hybrid scheme (solid lines)
and the pure LNV scheme (dashed lines) for diamond structure silicon.  
Results are shown for different cut-off radii: 3 hops (circles), 
5 hops (squares) and 7 hops (diamonds).  The radii are discussed in the text.}
\label{SiConv}
\end{figure}

The way in which the energy converges to
its ground-state value in the pure LNV scheme and in the LNV
stage of the hybrid scheme
is shown in Figures~\ref{CConv}
and~\ref{SiConv}, where we report the difference between the cohesive energy at
each iteration and the final ground-state cohesive energy, as a fraction of 
the final cohesive energy. The results show very clearly that the
McWeeny stage gets closer and closer to the correct ground state
as the cut-off radius increases. The rate of convergence in the
LNV stage is essentially the same as that found in the pure LNV scheme,
as expected. In all cases, the error in the energy decrease 
approximately exponentially with iteration number. 

\subsection{Carbon Vacancy}
When a vacancy is introduced into diamond-structure carbon,
the degeneracy of the dangling-bond states is broken
by a Jahn-Teller distortion which lowers the symmetry of the
system, and defect states appear in the band gap. This system
therefore gives us an interesting increase in complexity compared
with the perfect crystal.

Table~\ref{Tab:CVac} reports the numbers of iterations taken by the hybrid and 
pure LNV schemes.  The behaviour is similar to that found
for the perfect crystal, although for small cut-off radii the
total number of iterations in the hybrid scheme is now slightly
larger than for pure LNV. Nevertheless, the hybrid scheme is still
significantly faster than pure LNV in terms of CPU time.
For the present system, we have examined the consequences of
working at constant $\mu$ rather than constant electron number
in the LNV stage of the hybrid scheme. To do this, we have fixed $\mu$ at 
the value of ${\nabla N_o \cdot \nabla E / \nabla N_o \cdot \nabla N_o}$,
which is the correct definition of the chemical potential for electrons.  The 
final column of Table~\ref{Tab:CVac} shows the deviation of total electron
number from its nominal value of 252 (the calculation was done
with a system of 63 atoms) in the final ground state when the
calculation is done like this. The very small deviations,
which decrease with increasing cut-off radius, confirm our
expectation that $N_0$ automatically holds itself almost constant
in the LNV stage (see Sec.~\ref{RelMcWV}).

\subsection{Si(001) surface}

The Si(001) surface is a complex electronic system.  
Rebonding between surface atoms causes strong displacements
from perfect-lattice positions and the formation of dimers,
which themselves become buckled because of Jahn-Teller
distortion~\cite{Chadi}.
These effects give rise to bands of gap states.
Table~\ref{Tab:SiSurf}
shows the numbers of iterations required by the hybrid and pure LNV schemes. 
It is clear that 
in this case the hybrid scheme is significantly faster than pure LNV,
by a factor of between 3 and 5. We believe that the
slowness of pure LNV is a manifestation of ill conditioning
caused by gap states (see Sec.~\ref{VandMin}), which do not
appear to affect the McWeeny stage of the hybrid method.
As in the C vacancy case, we have examined the deviations of
electron number caused by running the LNV stage of the hybrid
method in constant-$\mu$ mode, and these are reported in the
final column of Table~\ref{Tab:SiSurf}. Once again, they
are very small, though not quite so small as in the vacancy case.

\subsection{Liquid silicon}
Liquid Si is a challenging system for any method, since it is
both disordered and metallic. But the important point here
is that dynamical simulation must be performed, so that energy conservation
is important, and this means that the consistency of
energy and ionic forces is crucial. We have pointed out in Sec.~\ref{RelMcWV}
that this consistency is ensured by the variational property of both
pure LNV and our hybrid method. By contrast, pure McWeeny is not
variational, so we are particularly interested to find out the
consequences of attempting to use pure McWeeny for this system.

The simulations were done using a repeating cell of 64 atoms at
an initial temperature of 3000~K. As before, the
orthogonal tight-binding model described in Ref.~\onlinecite{DRB}
was used. An important technical point for the pure LNV simulations 
is that the changes in bonding
from one step to the next are so strong that it is not helpful to use
the final density matrix from one step as the initial guess for
the density matrix at the next step\cite{LNVMD}. Instead,  
the standard initialisation
of $\frac{1}{2} {\bf I}$ is used.  It must be stressed that
because of the discontinuous changes of bonding caused by the
spatial cut-off of the density matrix, energy conservation
will be far from perfect whatever method is used. The point at issue
is therefore the relative quality of energy conservation
for different methods.

Another relevant technical point concerns the occupation numbers
to be used for the electronic states. There would be sound physical
arguments for using Fermi-Dirac occupation numbers corresponding
to the temperature of the simulation. For the present tests,
we have instead set the electronic temperature to zero, so that the
occupation numbers are exactly unity below the Fermi energy
and exactly zero above, following the theory of previous Sections.
The motivation for doing this is that it provides a more
stringent test of $O(N)$ methods for metallic systems, since increase
of the electronic temperature has the effect of localizing the
density matrix, as described in Ref.~\onlinecite{Bowler97}.

We have performed our {\em l}-Si simulations in all three
possible ways: pure LNV, hybrid, and pure McWeeny, using a spatial cut-off
of 3 hops on the density matrix.  We find
that the hybrid method is once again faster than pure LNV: on
average, the hybrid calculations require 13 McWeeny and 13 LNV iterations
per step, while pure LNV requires 26 iterations, so that
the hybrid method is between 25~\% and 50~\% faster
than pure LNV.

\begin{figure}
\begin{center}
\epsfxsize=80mm
\epsfbox{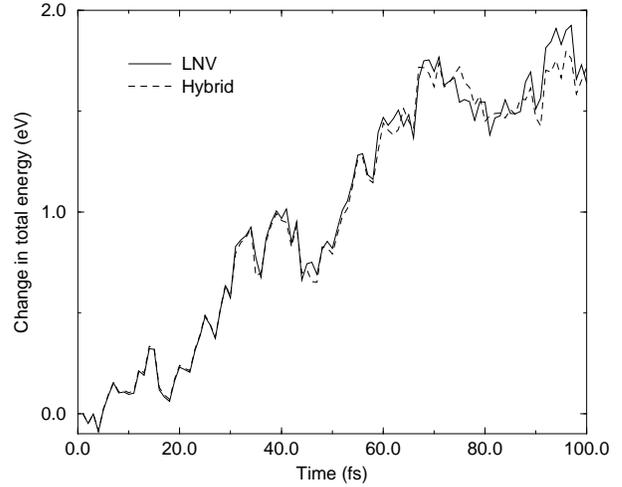}
\epsfxsize=80mm
\epsfbox{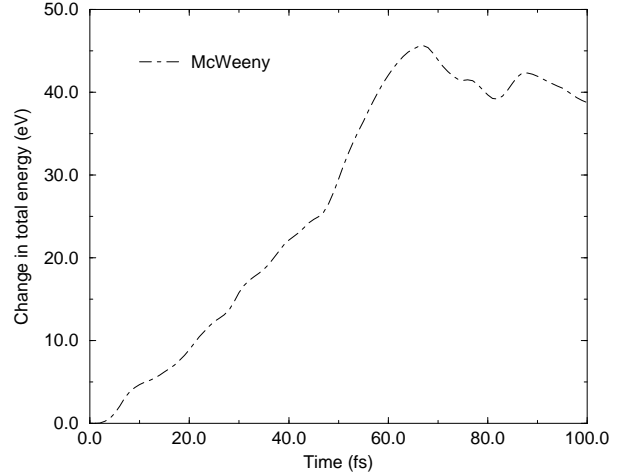}
\end{center}
\caption{Change in total energy (eV units) during molecular
dynamics runs on a 64-atom system of liquid Si for
(a) pure LNV (solid curve) and hybrid method (dashed curve)
and (b) pure McWeeny method.}
\label{fig:EConv}
\end{figure}

Fig.~\ref{fig:EConv} shows the results of our tests on energy conservation
over a period of 0.1~ps. The variation of the total energy
is almost exactly the same in the LNV and hybrid methods 
and consists of a steady upward drift of $\sim$~0.03~eV/atom
during the 0.1~ps period. This is already large, since it
corresponds to a temperature increase of {\em ca.}~100~K.
But with pure McWeeny, the increase of total energy is about 25 times
larger, and in our judgment this does not give a satisfactory
method of doing dynamics for this kind of system. We regard this
as a compelling reason for not using McWeeny by itself.

\section{Discussion}
\label{Disc}

The theoretical arguments presented in Secs.~\ref{McWV} and \ref{RelMcWV}
showed that the McWeeny and LNV techniques behave in very different
ways in the search for the ground state, and indicated that
a combination of the two techniques should be more efficient
and robust than either technique by itself. The practical
tests we have just reported fully support these ideas.
For the five systems we have studied, they confirm that
in terms of CPU time the hybrid technique is always faster,
and sometimes much faster, than the LNV technique by itself. They
also confirm that the hybrid technique is more robust, since
the LNV iterations used in the final stage show no sign of
the unstable behaviour that can occur if LNV is used thoughout.
We have seen that McWeeny used by itself, following the proposals
of Palser and Manolopoulos~\cite{Palser98}, can indeed produce
very accurate results for the ground-state energy, especially
when the spatial cut-off is increased, but that its lack
of consistency between energy and ionic forces can cause serious
problems in dynamical simulations. Importantly, our results also
demonstrate that the hybrid method allows one to work at
constant electron number with greater ease than when LNV is used
by itself.

Although we have chosen to work within orthogonal tight-binding
theory, the main ideas carry over directly to the non-orthogonal
case. Nunes and Vanderbilt~\cite{Nunes94} have already shown
how to generalize the LNV scheme to perform $O(N)$ tight-binding
calculations with non-orthogonal orbitals, and Palser and Manolopoulos
have shown that McWeeny purification can be generalized in the
same way. We demonstrate in Appendix~\ref{AppNO} that our two key
statements presented in Sec.~\ref{RelMcWV} remain valid in the
non-orthogonal case, provided the scalar product between matrices
is defined using the appropriate metric. The discussion in
Appendix~B indicates that the only significant problem in using
our hybrid $O(N)$ scheme for doing practical calculations
within non-orthogonal tight-binding theory is that one
needs the inverse of the overlap matrix $S_{i j}$ between orbitals.
However, we believe that an approximate inverse of $S_{i j}$ should suffice,
and ways of obtaining a suitable approximation have been proposed
by Palser and Manolopoulos~\cite{Palser98,InvS}.

The ideas we have presented should also be applicable to
$O(N)$ DFT. In a series of recent papers~\cite{hernandez95,%
hernandez96,goringe97,bowler98}
we have shown how a practical
$O(N)$ DFT scheme can be constructed, the guiding principle
being the minimization of the DFT total energy
with respect to the density matrix. The ideas have been implemented
in our code CONQUEST~\cite{goringe97}
({\em C}oncurrent $O(N)$ {\em QU}antum
{\em E}lectronic {\em S}tructure {\em T}echnique), in which
the density matrix is represented in a basis of localized
orbitals which are varied to minimize the total energy. In the
present form of CONQUEST, the minimization is performed in
three nested loops. In the innermost loop, the energy is minimized
wtih respect to the elements of the density matrix, and the
two outer loops have the tasks of achieving self-consistency
and minimizing with respect to the localized orbitals. The
operations performed in the innermost loop are therefore
identical to those performed in non-orthogonal
tight-binding theory, and the methods we have presented
here should be applicable without change to this part of
$O(N)$ DFT.

The relevance of the present ideas to $O(N)$ DFT gives another
reason why the LNV component of our hybrid scheme
is so important. In order to minimize the total energy
with respect to the localized orbitals, it is essential
to have an analytic expression for the gradient of the energy
with respect to variations of these orbitals. But the existence
of such an expression for the gradient relies crucially
on the energy being stationary  with respect to variations of the
density matrix. So once again the variational behaviour given
by LNV is essential, just as it is for calculating forces -- and
for basically the same reason. We are currently studying
the improvements that can be achieved in $O(N)$ DFT by applying
the ideas we have presented.

\section{Conclusions}
\label{Conc}

We have shown that a combination of McWeeny purification and
LNV minimization methods gives a robust and rapid means of searching for the
electronic ground state in the
framework of an $O(N)$ density-matrix scheme.  The McWeeny stage finds an 
idempotent density matrix quickly and efficiently, and the LNV stage then finds
the idempotent density matrix which minimises the total energy.  We have 
presented examples which demonstrate the advantages of this hybrid scheme, and 
shown why both stages are necessary. We have pointed out that
the main ideas can be generalized to non-orthogonal tight-binding
theory and $O(N)$ density functional theory.

\section*{Acknowledgments}
We are happy to acknowledge useful discussions with David Manolopoulos and
the early viewing of Ref.~\onlinecite{Palser98}.  We would also like to 
acknowledge useful discussions with Chris Goringe, and the use of his 
density matrix code, DensEl.
The work of DRB and the computer facilities used in the
calculations are supported by EPSRC grant GR/M01753. The work
of MJG is financially supported by CCLRC and GEC.

\appendix
\section{Proof of the two statements}
\label{AppOrtho}

We want to prove the two statements enunciated in Sec.~\ref{RelMcWV}: 
(1)~all McWeeny paths meet the idempotency surface orthogonally; 
(2)~for any point on the idempotency surface, the gradient of the LNV
function is tangential to the surface.

We show first how to characterise displacements within and
perpendicular to the tangent plane. Suppose $P$ is a point on
the idempotency surface and let $Q = I - P$, so that
$P^2 = P$ and $Q^2 = Q$. Consider displacements away from
$P$ represented by $P^\prime = P + \alpha \delta P$.
Any displacement $\delta P$ can be written as
\begin{equation}
\delta P = \delta P_\parallel + \delta P_\perp \;,
\end{equation}
where
\begin{eqnarray}
\delta P_\parallel &=& P \delta P Q + Q \delta P P \nonumber \\
\delta P_\perp     &=& P \delta P P + Q \delta P Q \; .
\end{eqnarray}
It is straightforward to show that any displacement of the
form $P^\prime = P + \alpha \delta P_\parallel$ is in
the tangent plane:
\begin{eqnarray}
P^{\prime \, ^2} &=& P^2 + \alpha ( P \delta P_\parallel +
\delta P_\parallel P ) + \alpha^2 \delta P_\parallel^2 \nonumber \\
              &=& P^\prime + \alpha^2 \delta P_\parallel^2 \; .
\end{eqnarray}
On the other hand, no displacement of the form $P^\prime =
P + \alpha \delta P_\perp$ can be in the tangent plane, since
a little algebra shows that
\begin{equation}
P^{\prime \, 2} = P^\prime + \alpha ( P \delta P P - Q \delta P Q ) +
\alpha^2 \delta P_\perp^2 \; .
\end{equation}
But the term linear in $\alpha$ cannot vanish unless $P \delta P P$
and $Q \delta P Q$ vanish separately, which implies that
$\delta P_\perp = 0$. It follows that the displaced point
$P^\prime = P + \alpha ( \delta P_\parallel + \delta P_\perp )$
is in the tangent plane if and only if $\delta P_\perp = 0$.

Now let $P$ be a point on the idempotency surface, and let
$P + \alpha \delta P$ be a point on a McWeeny path that leads
to $P$. Consider the matrix $P^\prime$ obtained by performing
a single purification on $P + \alpha \delta P$:
\begin{equation}
P^\prime = 3 ( P + \alpha \delta P )^2 -
2 ( P + \alpha \delta P )^3 \; .
\end{equation}
Because of the property of quadratic convergence, $P^\prime$
must deviate from $P$ only by a term of order $\alpha^2$.
The condition that the linear term in $P^\prime$ vanishes is:
\begin{equation}
\delta P P - 2 P \delta P P + P \delta P = 0 \; .
\end{equation}
If we decompose $\delta P$ into its tangential and perpendicular
components by writing $\delta P = \delta P_\parallel +
\delta P_\perp$, then this condition becomes:
\begin{equation}
\delta P_\parallel = 0 \; .
\end{equation}
The implication is that any McWeeny path leading to $P$
must approach along a direction of the form $\delta P_\perp$.
Since $( \delta P_\parallel , \delta P_\perp ) = 0$,
this proves that all McWeeny paths meet the idempotency
surface at right angles to the surface.

We now prove that the gradient of the LNV
function of eqn~(\ref{eq:GP}) is
tangential to the surface. We define the elements
$F_{i j}$ of this (negative) gradient as:
\begin{equation}
F_{i j} = - \partial \Omega / \partial {\tilde{\rho}}_{j i} \; ,
\end{equation}
so that from the definition given in eqn~(\ref{eq:GP}) we have:
\begin{equation}
F_{i j} = 3 ( H^\prime \tilde{\rho} + \tilde{\rho} H^\prime )_{i j} -
2 ( {\tilde{\rho}}^2 H^\prime + \tilde{\rho} H^\prime \tilde{\rho} +
H^\prime {\tilde{\rho}}^2 )_{i j} \; ,
\end{equation}
where $H^\prime \equiv H - \mu I$. If $\tilde{\rho}$ is a point $P$ on
the idempotency surface, then:
\begin{equation}
F = H^\prime P - 2 P H^\prime P +
P H^\prime \; .
\end{equation}
It is now straightforward to show that $P^\prime =
P + \alpha F$ is a tangent line:
\begin{eqnarray}
P^{\prime \, 2} & = & P^2 + \alpha ( P F +
F P ) + \alpha^2 F^2 \nonumber \\
& = & P + \alpha ( H^\prime P -
2 P H^\prime P + P H^\prime ) +
\alpha^2 F^2 \nonumber \\
& = & P^\prime + \alpha^2 F^2 \; .
\end{eqnarray}

\section{The non-orthonormal case}
\label{AppNO}

We show here that the two statements enunciated in Sec.~\ref{RelMcWV}
remain true when we generalize to non-orthonormal
basis functions. In this case,
the energy eigenvalues $\epsilon_n$ are determined by
the generalized eigenvalue equation:
\begin{equation}
\sum_{j=1}^N {\bar{H}}_{i j} {\bar{c}}_{j n} =
\epsilon_n \sum_{j=1}^N {\bar{S}}_{i j} {\bar{c}}_{j n} \; ,
\end{equation}
where ${\bar{S}}_{i j}$ is the overlap matrix. (Our notation
gives an over-bar to quantities in the
non-orthonormal case to distinguish them from the corresponding
quantities in the orthonormal case.) With $N_0$ occupied
states, the ground-state energy $E_0$ is then determined by:
\begin{equation}
E_0 = \min {\rm Tr} \, ( \bar{H} \bar{\rho} ) \; ,
\end{equation}
subject to the constraints that $\bar{\rho}$ is a `generalized
projector':
\begin{equation}
\bar{\rho} \bar{S} \bar{\rho} = \bar{\rho}
\label{eq:GenProj}
\end{equation}
and that ${\rm Tr} \, ( \bar{S} \bar{\rho} ) = N_0$. If we work at
constant chemical potential, then we minimize the
grand potential $\Omega$:
\begin{equation}
\Omega_0 = \min {\rm Tr} \, ( \bar{H} - \mu \bar{S} ) \bar{\rho}
\end{equation}
subject to $\bar{\rho}$ being a generalized projector. This whole
scheme corresponds precisely to the orthonormal scheme through
the relations:
\begin{eqnarray}
\bar{H} & = & {\bar{S}}^{1/2} H {\bar{S}}^{1/2} \nonumber \\
\bar{\rho} & = & {\bar{S}}^{- 1/2} \rho {\bar{S}}^{- 1/2} \; .
\end{eqnarray}

From this correspondence, it is clear that the McWeeny purification
of the density matrix $\rho$ is accomplished by iteration
of the algorithm:
\begin{equation}
\rho = 3 \tilde{\rho} S \tilde{\rho}
- 2 \tilde{\rho} S
\tilde{\rho} S \tilde{\rho} \; .
\end{equation}
(To simplify the notation, we now drop the over-bars.)
The scheme of Palser and Manolopoulos~\cite{Palser98} 
can be implemented by starting from an initial density matrix given by:
\begin{equation}
\rho^{(0)} = \frac{1}{2} \xi ( \mu S^{-1} -
S^{-1} H S^{-1} ) + \frac{1}{2}
S^{-1}
\end{equation}
as has already been pointed out in Ref.~\onlinecite{Palser98}. 
The constant-$N_0$ algorithm can also be recast in non-orthonormal form.

Similarly, the Nunes and Vanderbilt (NV) method~\cite{Nunes94} 
at constant $\mu$ consists of minimizing $\Omega$ given by:
\begin{equation}
\Omega = {\rm Tr} \, ( H - \mu S )
( 3 \tilde{\rho} S \tilde{\rho} -
2 \tilde{\rho} S \tilde{\rho} S
\tilde{\rho} ) \; .
\label{eq:NOGP}
\end{equation}
This formulation of the NV method for the non-orthonormal
case has already been discussed in 
Refs.~\onlinecite{Nunes94,hernandez95}.

The two key statements of Sec.~\ref{RelMcWV} remain valid provided
we use the appropriate metric for defining scalar products of matrices.
Instead of eqn~(\ref{eq:AB}), we must use the definition:
\begin{equation}
( A , B ) = {\rm Tr} \, ( A^\dagger S B S ) \; ,
\label{eq:NOMet}
\end{equation}
which means that the matrices $A$ and $B$ are orthogonal
if ${\rm Tr} \, ( A^\dagger S B S ) = 0$. The `idempotency
surface' must, of course, the taken to mean the manifold
of all matrices satsifying eqn~(\ref{eq:GenProj}). The methods of 
Appendix~\ref{AppOrtho}
are then straightforwardly repeated to
show that McWeeny paths meet the idempotency surface
orthogonally.

We now turn to the LNV gradient. Naively, one might
be inclined to work with the (negative) gradient
$F_{i j} = - \partial \Omega / \partial {\tilde{\rho}}_{j i}$
as defined in the orthogonal case. However, as has been
stressed recently by White {\em et al.}~\cite{White97}, for a metric defined
by eqn.~(\ref{eq:NOMet}), it would
not be tensorially correct to update the
density matrix $\tilde{\rho}$ using a gradient
defined in this way, since transformations of the basis
make $\tilde{\rho}_{i j}$
behave as a contravariant quantity but the derivative
$F_{i j}$ as a covariant quantity. Instead, one should work with the
contravariant gradient $\Phi$ defined as:
\begin{equation}
\Phi_{i j} = ( S^{-1} F S^{-1})_{i j} \; .
\end{equation}
If one were to work with the metric defined by eqn.~(\ref{eq:AB}), then 
the original gradient of $F_{ij}$ is correct (though the two statements in
Section~\ref{RelMcWV} {\it require} the metric defined by
eqn.~\ref{eq:NOMet}).

It is straightforward to show that $\Phi$ is tangential to
the idempotency surface. To do this, note from eqn~(\ref{eq:NOGP}) that the
gradient $F$ of $\Omega$ is given by:
\begin{eqnarray}
F_{i j} & = & - \partial \Omega / \partial {\tilde{\rho}}_{j i}
\nonumber \\
& = & 3 ( S \tilde{\rho} H^\prime + H^\prime \tilde{\rho} S )_{i j}
\nonumber\\
& - & 2 ( S \tilde{\rho} S \tilde{\rho} H^\prime +
S \tilde{\rho} H^\prime \tilde{\rho} S +
H^\prime \tilde{\rho} S \tilde{\rho} S )_{i j} \; ,
\end{eqnarray}
where $H^\prime \equiv H - \mu S$. If $\tilde{\rho}$ is a
point $P$ on the idempotency surface, then this reduces to:
\begin{equation}
F = S P H^\prime -
2 S P H^\prime P S + H^\prime P S \; ,
\end{equation}
and the contravariant gradient is:
\begin{equation}
\Phi = P H^\prime S^{-1} -
2 P H^\prime P + S^{-1} H^\prime P \; .
\end{equation}
It then follows that $P^\prime \equiv P + \alpha \Phi$ is in
the tangent plane:
\begin{eqnarray}
P^\prime S P^\prime & = & P + \alpha ( \Phi S P + P S \Phi ) +
\alpha^2 \Phi S \Phi \nonumber \\
& = &  P^\prime + \alpha^2 \Phi S \Phi \; .
\end{eqnarray}

\begin{table}
\begin{center}
\begin{tabular}{|c|cc|c|c|}
Radius (hops) & \multicolumn{2}{c|}{Iterations (Hybrid)} & Iterations (LNV) & Norm \\
 & McWeeny & LNV & & \\
\hline
3 & 9  & 8 & 17 & 0.049 \\
5 & 10 & 7 & 17 & 0.015 \\
7 & 11 & 5 & 16 & 0.007 \\
\end{tabular}
\end{center}
\caption{Aspects of the convergence to the ground state for the hybrid scheme
and a pure LNV scheme for diamond structure carbon.  The first column shows
the spatial cut-off applied to the density matrix.  The second and third show  
the number of iterations required in the McWeeny and LNV stages of the hybrid 
scheme, and the fourth the number of iterations required by a pure LNV scheme.
The last column shows the norm of the difference between the final McWeeny 
density matrix and the final density matrix (see Eq.~(\protect\ref{eq:Norm})
in Section~\protect\ref{RelMcWV} for the definition of norm).  The convergence
criterion was a fractional difference of 10$^{-8}$ between the cohesive 
energies in successive line minimisations.}
\label{Tab:CConv}
\end{table}

\begin{table}
\begin{center}
\begin{tabular}{|c|cc|c|c|}
Radius (hops) & \multicolumn{2}{c|}{Iterations (Hybrid)} & Iterations (LNV) & Norm \\
 & McWeeny & LNV & & \\
\hline
3 & 11 & 10 & 21 & 0.078 \\
5 & 12 & 10 & 21 & 0.043 \\
7 & 13 &  9 & 22 & 0.024 \\
\end{tabular}
\end{center}
\caption{Aspects of the convergence to the ground state for the hybrid scheme
and a pure LNV scheme for diamond structure silicon.  Results are displayed 
as in Table~\protect\ref{Tab:CConv}.}
\label{Tab:SiConv}
\end{table}

\begin{table}
\begin{center}
\begin{tabular}{|c|cc|c|c|}
Radius (hops) & \multicolumn{2}{c|}{Iterations (Hybrid)} & Iterations (LNV) & $\Delta N_o$ \\
 & McWeeny & LNV & & \\
\hline
3 & 10 & 9 & 16 & 0.015  \\
5 & 11 & 8 & 18 & 0.0006 \\
7 & 12 & 6 & 18 & 0.0004 \\
\end{tabular}
\end{center}
\caption{Aspects of the convergence to the ground state for the hybrid scheme
and a pure LNV scheme for a relaxed vacancy in carbon.  The first four columns
are as in Table~\protect\ref{Tab:CConv}, while the last column shows the 
deviation, $\Delta N_o$, from the total electron number (252) after the 
LNV stage of the hybrid minimisation.}
\label{Tab:CVac}
\end{table}

\begin{table}
\begin{center}
\begin{tabular}{|c|cc|c|c|}
Radius (hops) & \multicolumn{2}{c|}{Iterations (Hybrid)} & Iterations (LNV) & $\Delta N_o$ \\
 & McWeeny & LNV & & \\
\hline
3 & 15 & 14 & 73 & 0.0311 \\
5 & 16 & 13 & 79 & 0.0213 \\
7 & 17 & 10 & 82 & 0.0221 \\
\end{tabular}
\end{center}
\caption{Aspects of the convergence to the ground state for the hybrid scheme
and a pure LNV scheme for the Si(001) surface. The first four columns
are as in Table~\protect\ref{Tab:CConv}, while the last column shows the 
deviation, $\Delta N_o$, from the total electron number (168) after the 
LNV stage of the hybrid minimisation.}
\label{Tab:SiSurf}
\end{table}

\end{document}